\documentclass[fleqn,usenatbib]{mnras}

\usepackage{newtxtext,newtxmath}
\usepackage[T1]{fontenc}

\DeclareRobustCommand{\VAN}[3]{#2}
\let\VANthebibliography\thebibliography
\def\thebibliography{\DeclareRobustCommand{\VAN}[3]{##3}\VANthebibliography}

\usepackage{graphicx}	% Including figure files
\usepackage{amsmath}	% Advanced maths commands

\title[RBH-1 cooling and entrainment]{A Dynamical Test for Cooling-Induced Entrainment in a Runaway Supermassive Black Hole Tail}

\author[I. Kaul et al.]{
Ish Kaul\thanks{E-mail: ikaul@ucsb.edu} and 
S. Peng Oh
\\
Department of Physics, University of California, Santa Barbara, CA 93106, USA\\
}

\date{Accepted XXX. Received YYY; in original form ZZZ}

\pubyear{\the\year{}}

\begin{document}
\label{firstpage}
\pagerange{\pageref{firstpage}--\pageref{lastpage}}
\maketitle

\begin{abstract}
Radiative turbulent mixing layers are widely invoked to explain the survival, growth, and entrainment of cold gas in hot astrophysical flows, but quantitative dynamical tests have remained scarce. RBH-1, the first confirmed runaway supermassive black hole, offers a rare opportunity to test this framework: JWST observations show a $62\,\mathrm{kpc}$ tail of cold H$\alpha$ and [O III]-emitting gas behind a source moving at $\sim 950\,\mathrm{km\,s^{-1}}$ through the hot circumgalactic medium, with a coherent velocity gradient of $\sim 200\,\mathrm{km\,s^{-1}}$ along the tail. Using 3D hydrodynamical simulations together with turbulent mixing-layer theory, we model the coherent downstream tail. We find that the observed downstream deceleration is well reproduced by accretion-induced drag from radiative mixing layers, and that without radiative cooling no coherent cold tail forms. We also derive a direct connection between the tail deceleration and the cooling luminosity, yielding predictions for future measurements of the cooling luminosity profile. RBH-1 therefore provides a rare quantitative dynamical stress test of radiative mixing-layer physics in an astrophysical system.

%%%Recently, JWST observations confirmed the first runaway supermassive black hole (RBH-1) moving at $\sim 950 \ \rm km \ s^{-1}$ through the ambient hot circumgalactic medium (CGM). This RBH-1 left behind a 62 kpc long trail of cold H$\alpha$ and [OIII] emitting gas in its wake of radius 0.7 kpc and a velocity gradient of $\sim 200 \ \rm km\ s^{-1}$ along its length. Using 3D hydrodynamical simulations, along with theoretical predictions from turbulent mixing layer theory, we model the coherent downstream wake, leaving the more complex head where cold gas is initially seeded for future work. We find that the observed velocity gradient is naturally and quantitatively explained by accretion induced drag from radiative mixing layers. We also use the observed deceleration to make direct predictions for the surface brightness of cooling radiation, which could be probed in future observations. These results provide one of the strongest quantitative dynamical cases to date for cooling-induced entrainment in an astrophysical system. 
\end{abstract}


%%%%%%%%%%%%%%%%%%%%%%%%%%%%%%%%%%%%%%%%%%%%%%%%%%

%%%%%%%%%%%%%%%%% BODY OF PAPER %%%%%%%%%%%%%%%%%%

\section{Introduction}
Radiative turbulent mixing layers are widely invoked to explain the survival, growth, and emission of cold gas embedded in hot astrophysical flows, from galactic winds to the circumgalactic medium (CGM). When cold (\(\sim 10^4\,\mathrm{K}\)) gas interacts with a hot (\(\sim 10^6\,\mathrm{K}\)) background flow, shear at the interface drives mixing, producing intermediate-temperature gas that can cool efficiently and join the cold phase \citep{begelman90,kwak10,ji19,fielding2020,Tan2021,chen2023,zhao23,sharma25,marin-gilabert2025}. If cooling is sufficiently rapid, this allows cold structures such as clouds \citep{Gronke2018} and cold streams \citep{mandelker2020} to survive and even grow despite disruptive hydrodynamic instabilities. This emerging paradigm has been extensively explored in the literature, with additional physics such as magnetic fields \citep{gronke2020, hidalgo-pineda2024}, thermal conduction \citep{jennings2023}, viscosity \citep{Li2020}, dust \citep{Chen2024}, and gravity \citep{Tan2023, Kaul2025} shown to influence the dynamics, morphology, and emission properties of the gas; see \citet{faucher-giguere2023} for a review. Yet despite this theoretical progress, quantitative dynamical tests of radiative mixing-layer physics in real astrophysical systems remain scarce.

A particularly direct prediction of this framework is that growth of the cold phase through mixing and cooling should also transfer momentum. Recent work has shown that when hot gas is entrained, mixed, and radiatively cooled into a moving cold structure, momentum conservation produces an accretion-induced drag force whose strength is set by the cold-gas growth time \citep{Tan2023,Kaul2025}. This idea has so far been developed mainly for infalling systems such as high-velocity clouds and cluster filaments \citep{Fox2005,Westmeier2018,Lim2008}, but the same physics should apply more generally to externally forced or wind-driven cold gas. If radiative mixing layers regulate both mass growth and deceleration, then the velocity structure of a cold tail provides a direct observational test of the theory.

RBH-1 offers a rare opportunity to perform such a test. The first compelling candidate for a runaway supermassive black hole (SMBH) was identified by \citet{vanDokkum2023}, who reported a \(\sim 62\,\mathrm{kpc}\) linear feature extending from a \(z=0.96\) galaxy, with a bright [O III] knot at its tip. Follow-up JWST/NIRSpec and HST IFU observations by \citet{vanDokkum2026} strongly supported the runaway-SMBH interpretation, showing that the steep \(\sim 600\,\mathrm{km\,s^{-1}}\) velocity gradient across the shocked head is consistent with a bow shock driven by an SMBH moving at \(\sim 950\,\mathrm{km\,s^{-1}}\) through the ambient CGM at inclination \(i=29^\circ\). Beyond the shocked head, extended H\(\alpha\) and [O III] emission traces a long, cold tail with a coherent \(\sim 200\,\mathrm{km\,s^{-1}}\) velocity gradient along its length. This combination of a hot ambient medium, a long-lived cold tail, and spatially resolved kinematics makes RBH-1 an unusually clean laboratory for testing cooling-induced entrainment.

In this Letter, we ask whether the coherent downstream RBH-1 tail can be understood as a dynamical consequence of radiative mixing layers. We do not attempt to explain the detailed origin of the cold gas or the more complex near-head transition region; instead, we focus on the downstream coherent tail, where the cold gas decelerates steadily. We combine analytic estimates based on mixing-layer theory with 3D hydrodynamical simulations including radiative cooling to test whether the observed velocity gradient is consistent with accretion-induced drag, and to derive a connection between the tail deceleration and the cooling luminosity that can be tested with future observations.

\section{Analytic Estimates}
\label{sec:analytic}

\subsection{Dynamics from mixing layers}

Conventional ram-pressure drag is far too weak to explain the RBH-1 tail deceleration: for \(\chi\sim100\), \(r_{\rm cl}\sim1\,\mathrm{kpc}\), and \(v\sim900\,\mathrm{km\,s^{-1}}\), one finds \(t_{\rm drag,ram}\sim 2.7\chi r_{\rm cl}/v \sim 270\,\mathrm{Myr}\) for the initial cloud and \(t_{\rm drag,ram}\sim 2\chi L/v \sim 8\,\mathrm{Gyr}\) for a \(L\sim40\,\mathrm{kpc}\) tail, both much longer than the \(\sim70\,\mathrm{Myr}\) system lifetime. We therefore consider braking by growth of the cold phase through mixing and radiative cooling.

We now consider the dynamics of a cold structure undergoing accretion-induced drag. The momentum equation is
\begin{equation}
    \frac{d}{dt}(mv)=\dot m\,v_{\rm src},
\end{equation}
where \(m\) and \(v\) are the cold-gas mass and velocity, and \(v_{\rm src}\) is the bulk velocity of the newly added material as it joins the cold phase. Writing the mass growth in terms of the growth time \(t_{\rm grow}\equiv m/\dot m\), we obtain
\begin{equation}
    \dot v = -\frac{v-v_{\rm src}}{t_{\rm grow}}.
\end{equation}
As shown in \citet{Tan2023}, growth of the cold phase through mixing and cooling therefore produces a braking force. For simplicity, we parameterize the source velocity as a fraction of the cold-gas speed,
\begin{equation}
    v_{\rm src}=\alpha v,
\end{equation}
where \(0\le \alpha <1\). We expect $\alpha \ll 1$ for a CGM with little bulk motion. This gives
\begin{equation}
    \dot v = -(1-\alpha)\frac{v}{t_{\rm grow}}.
\end{equation}
The \(\alpha\rightarrow 0\) limit recovers the form used in \citet{Tan2023}. To compare with observations, we convert the time dependence to a spatial dependence along the tail:
\begin{equation}
    \frac{dv}{dt}=v\frac{dv}{dx}=-(1-\alpha)\frac{v}{t_{\rm grow}}
    \;\;\Longrightarrow\;\;
    \frac{dv}{dx}=-\frac{1-\alpha}{t_{\rm grow}}.
\end{equation}
For intuition, if \(t_{\rm grow}\) and \(\alpha\) were constant, this integrates to
\begin{equation}
\label{eqn:velocity_equation}
    v(x)=v_0-\frac{(1-\alpha)x}{t_{\rm grow}}.
\end{equation}
In practice, however, \(t_{\rm grow}\) varies with position through the delayed onset of mixing and the evolution of the cold phase. Using analytical scalings calibrated to radiative turbulent mixing-layer simulations, \citet{Tan2023} found
\begin{align}
\label{eqn:tgrow_equation}
        t_{\rm grow} = \frac{35 \mathrm{Myr}}{w_{\rm kh}(x)}\left(\frac{f_A}{0.23} \right)\left(\frac{c_{\rm s, hot}}{150 \mathrm{km\ s^{-1}}}\right)^{3/5}\left(\frac{\chi}{100}\right)\left(\frac{r_{\rm cl}}{100 \ \rm pc}\right)^{3/4} \nonumber\\
    \times \left(\frac{t_{\rm cool}}{0.03 \ \rm Myr}\right)^{1/4}\left(\frac{m}{m_0}\right)^{1/6}
\end{align}
where $w_{\rm kh}(x) = \mathrm{min}\left(1,\frac{x}{f_{\rm kh}r_{\rm cl}}\right)$ is a factor that accounts for the delayed onset of turbulence and mixing, $f_A$ is an order unity fudge factor, $c_{\rm s,hot}$ is the sound speed of the hot gas, $\chi$ is the overdensity of the cold cloud, $r_{\rm cl}$ is the cloud radius, $t_{\rm cool}$ is the minimum cooling time, $m/m_0$ is the cold gas mass fraction. Note that this is the growth time in the supersonic case, where the turbulent velocity saturates and does not depend on the shear velocity \citep{Yang2023}. 

\subsection{Dynamics from observables}

How can future observations test the connection between tail deceleration and radiative cooling? For a steady tail, momentum conservation gives
\begin{equation}
\label{eqn:momentum_flux}
    \dot M_{\rm tail}(x)\frac{dv}{dx}=-(v-v_{\rm src})S_{\rm rad}^{\rm add},
\end{equation}
where \(\dot M_{\rm tail}(x)=\lambda(x)v(x)\) is the local cold-gas mass flux along the tail, \(v_{\rm src}\) is the bulk velocity of the newly added material as it joins the cold phase, and \(S_{\rm rad}^{\rm add}\) is the local mass-addition rate per unit length. We neglect the corresponding loss term, \(-v_{\rm loss}S_{\rm rad}^{\rm loss}\), because in our simulations \(v_{\rm loss}\sim v_{\rm tail}\), so material lost from the cold phase carries nearly the same specific momentum as the tail and contributes negligibly to its braking.

We write the cooling source term as
\begin{equation}
    S_{\rm rad}^{\rm add}=\frac{d\dot M_{\rm rad}}{dx}
    =\frac{1}{h_{\rm eff}}\frac{dL_{\rm bol}}{dx},
\end{equation}
where
\begin{equation}
    h_{\rm eff}=\frac{5}{2}\frac{k_B T_{\rm hot}}{\mu m_p}+\frac{1}{2}(v-v_{\rm src})^2
\end{equation}
is an effective specific enthalpy that includes both the thermal enthalpy of the hot gas and the relative bulk kinetic energy of the joining material. In practice, we find that the kinetic term is subdominant in our simulations, so \(h_{\rm eff}\) is set mainly by the hot-gas enthalpy. The dominant uncertainty in this conversion is therefore only at the level of a factor of a few (\(\sim 3\)), and mainly affects the normalization of the inferred \(t_{\rm grow}\).

Substituting \(\dot M_{\rm tail}(x)=\lambda(x)v(x)\) into Equation~\ref{eqn:momentum_flux}, we obtain
\begin{equation}
\label{eqn:v_obs_drag}
    \frac{dv}{dx}
    =-\frac{1-v_{\rm src}/v}{\lambda(x)h_{\rm eff}}\frac{dL_{\rm bol}}{dx}.
\end{equation}
Writing the source velocity as \(v_{\rm src}=\alpha v\) gives
\begin{equation}
\label{eqn:v_obs_pred}
    v_{\rm analytic}
    =v_0-(1-\alpha)\int \frac{dL_{\rm bol}/dx}{h_{\rm eff}\lambda(x)}\,dx.
\end{equation}
Equation~\ref{eqn:v_obs_pred} is therefore the same braking law as Equation~\ref{eqn:velocity_equation}, but with the local growth time inferred from observable quantities rather than from mixing-layer theory:
\begin{equation}
    t_{\rm grow}^{\rm obs}(x)\equiv \frac{\lambda(x)h_{\rm eff}}{dL_{\rm bol}/dx}.
\end{equation}

In practice, this test requires two observational inputs. First, the bolometric cooling profile \(dL_{\rm bol}/dx\) can be estimated from spatially resolved emission-line measurements along the tail, together with an ionization/cooling model to convert the observed line luminosity into a total cooling rate. Second, the linear mass density \(\lambda(x)=\int \rho_{\rm cold}\,dA\) can be estimated from the observed tail width and cold-gas density, for example using density-sensitive line diagnostics or an emission-measure estimate combined with an assumed path length and filling factor. Spatially resolved spectroscopy of the RBH-1 tail could therefore test whether the observed deceleration is quantitatively consistent with cooling-induced mass loading.

As shown below, the luminosity-based and mixing-layer closures give similar \(t_{\rm grow}\) in the simulations. In particular, over much of the coherent tail both \(dL_{\rm bol}/dx\) and \(\lambda(x)\) are approximately constant, so this closure implies an approximately constant local \(t_{\rm grow}\).

\section{Simulation Setup}
\label{sec:numerics}

We perform 3D hydrodynamical simulations using the publicly available \texttt{Athena++} code \citep{Stone2020}. All runs use a regular Cartesian grid, the HLLC Riemann solver, piecewise-linear reconstruction applied to primitive variables, and third-order Runge--Kutta time integration. Our fiducial domain has dimensions \(36\,\mathrm{kpc}\) in the downstream \(x\)-direction and \(6\,\mathrm{kpc}\) in both transverse directions, resolved by \(1536\times256\times256\) cells. We impose outflow boundary conditions on the transverse boundaries and the right \(x\)-boundary, and an inflowing wind on the left \(x\)-boundary.

The simulations are designed to model only the coherent downstream RBH-1 tail. We therefore begin the computational domain \(12\,\mathrm{kpc}\) downstream of the RBH-1 head, excluding the observed near-head region where the velocity profile remains approximately flat. At the start of the modeled region, the observed projected tail speed is \(\sim 350\,\mathrm{km\,s^{-1}}\); adopting the inclination \(i=29^\circ\) inferred by \citet{vanDokkum2026} gives a deprojected wind speed of \(v=350\,\mathrm{km\,s^{-1}}/\sin i \approx 700\,\mathrm{km\,s^{-1}}\). The background medium is initialized as uniform hot gas with \(T=10^6\,\mathrm{K}\) and \(n_{\rm H}=5\times10^{-3}\,\mathrm{cm^{-3}}\). Unlike standard cloud-crushing simulations, we do not apply frame tracking, since the domain is large enough to follow the full \(\sim 35\,\mathrm{kpc}\) coherent tail.

\textit{Brinkman Penalized Sphere Initialization.}
The observed RBH-1 wake appears to contain two distinct dynamical regimes: an initial \(\sim 12\,\mathrm{kpc}\) near-head transition region, where the cold phase is first produced and the velocity profile remains approximately flat, and a downstream coherent tail, where the cold gas decelerates steadily. This Letter is concerned only with the latter regime. A full model of the near-head transition would require additional assumptions about the origin and launch of the cold gas, which remain uncertain and are deferred to future work. Our setup is therefore intended as a minimal model of a continually fed wake, not a literal model of the RBH-1 head. To provide a compact, persistent source that remains in the domain and feeds the wake without prescribing a cold-stream injection by hand, we implement a rigid core using Brinkman penalization \citep{Liu2007}, a well-established method for modeling solid obstacles in engineering flows. We damp the velocity within the core on a timescale shorter than any physical timescale in the simulation, and assign it a density \(10^6\) times the ambient value, so that it remains effectively stationary and dynamically decoupled from the wind. We initialize a shell of cold \(10^4\,\mathrm{K}\) gas around the core to represent cold gas already present at the observed onset of the coherent downstream tail, and then ask whether its subsequent deceleration can be explained by cooling-induced mass loading. The immediate braking seen just downstream of the Brinkman sphere is therefore a consequence of this initialization, and should not be interpreted as a model for the observed near-head region of RBH-1, where the velocity profile remains approximately flat.

\textit{Cooling.}
We include radiative cooling assuming collisional ionization equilibrium and solar metallicity, using the \citet{Gnat2007} cooling table. To mimic background heating and prevent the ambient CGM from cooling, we suppress cooling above \(6\times10^5\,\mathrm{K}\). Throughout the paper, we define ``cold'' gas as material with \(10^4\,\mathrm{K}<T<3\times10^4\,\mathrm{K}\).

Table~1 summarizes the simulation suite. Here \(r_{\rm core}\) is the Brinkman-core radius, \(n_{\rm bkg}\) is the ambient hydrogen number density, and \(l_{\rm shell}\) is the thickness of the initial cold shell. In the run names, the number following ``c'' denotes the ratio \(r_{\rm core}/l_{\rm shell}\). The adiabatic control run and the single aligned-field test discussed below are auxiliary runs and are not listed in Table~1.
\begin{table}
	\centering
	\caption{List of Simulations}
	\label{tab:listOfSims}
	\begin{tabular}{lcccr} 
		\hline
            \hline
        Name & $r_{\rm core}$\ (kpc) & $n_{\rm bkg}$\ (cm$^{-3}$)  & $l_{\rm shell}$\ (kpc)  & $ \Delta x$ \\
        \hline
        \texttt{r0.8\_c1.1}& 0.8 & $5\times10^{-3}$ & 0.08 & 23pc\\
        \texttt{r0.8\_c1.1\_n3e-3}& 0.8 & $3\times10^{-3}$ & 0.08 & 23pc\\
        
        \texttt{r0.8\_c1.5}& 0.8 & $5\times10^{-3}$ & 0.4 & 23pc\\
        
        \texttt{res} & 0.8 & $5\times10^{-3}$ & 0.4 & 12pc\\
        \texttt{r0.8\_c2}& 0.8 & $5\times10^{-3}$ & 0.8 & 23pc\\
        \texttt{r2\_c2}& 2 & $5\times10^{-3}$ & 2 & 23pc\\

        \hline
        \hline
	\end{tabular}
	\\ \ \\
    \label{tab:sims}
%\caption{List of simulations.}
\end{table}

\section{Results}
\label{sec:results}

We now present the results of simulations that vary the shell thickness and Brinkman-core radius. All runs are evolved to at least the inferred RBH-1 lifetime, \(t\sim 70\,\mathrm{Myr}\) \citep{vanDokkum2026}.

\begin{figure*}
    \includegraphics[width=0.495\textwidth]{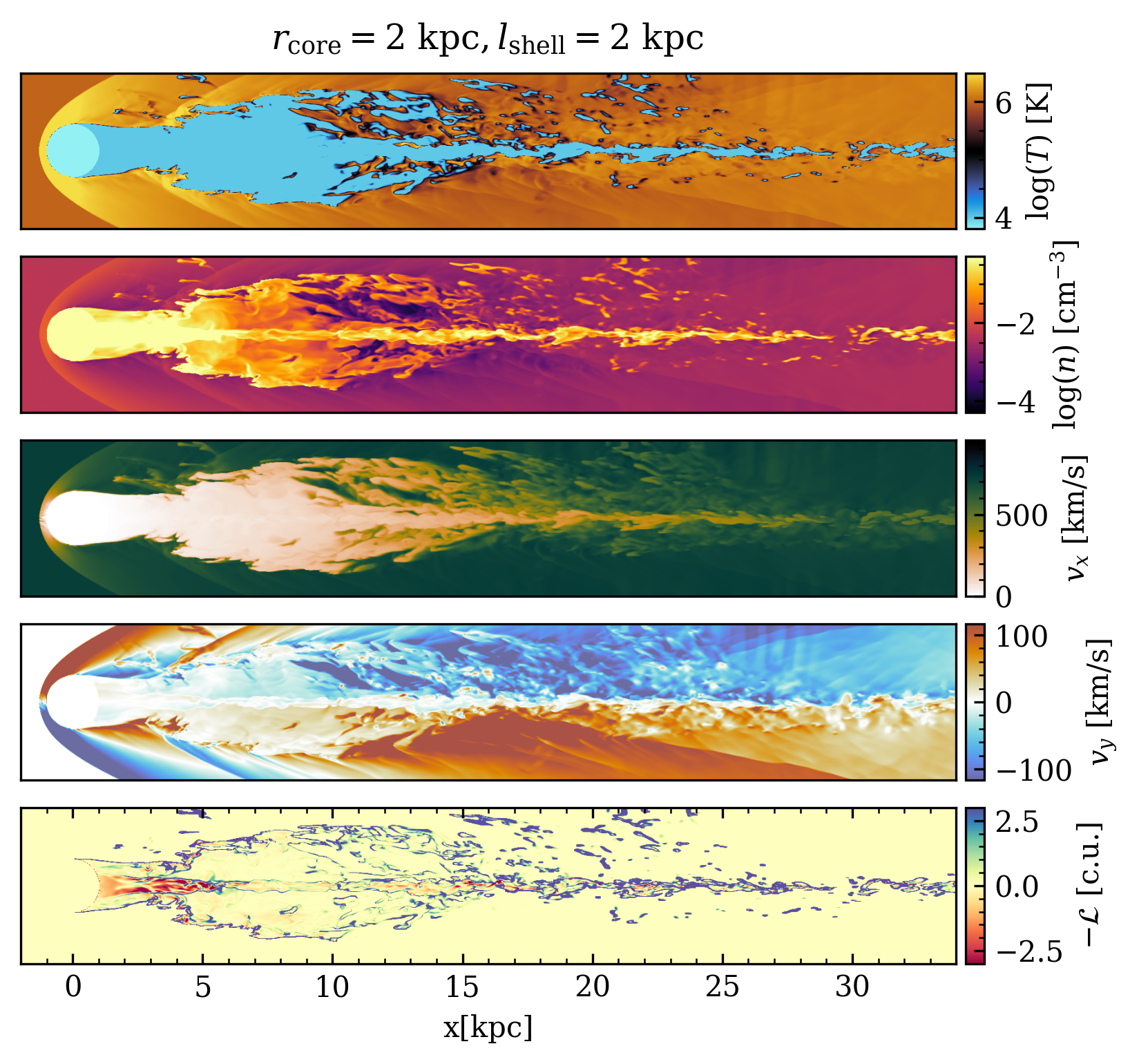}
    \includegraphics[width=0.495\textwidth]{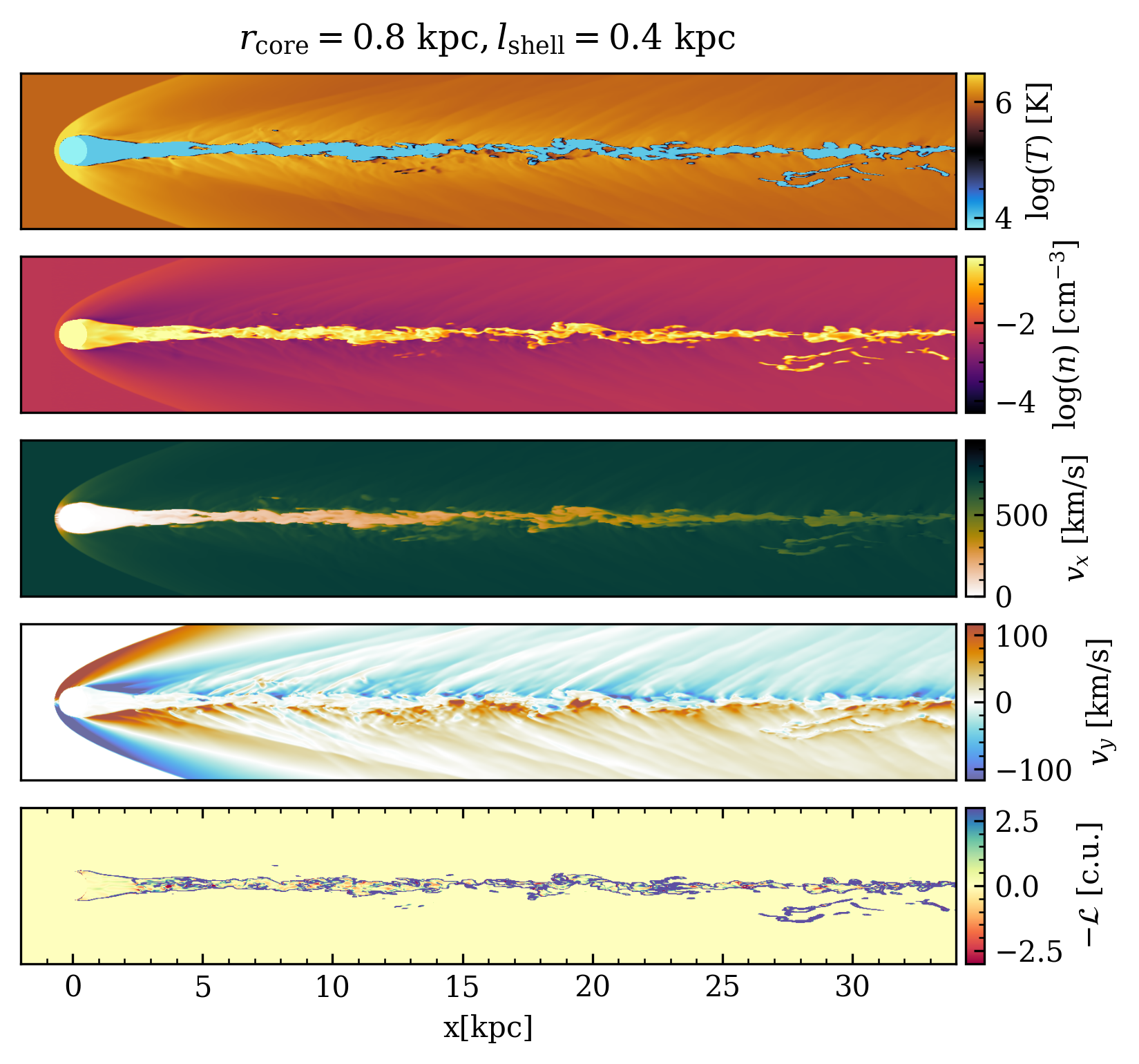}
    \caption{Snapshots of temperature, number density, velocity, and cooling luminosity for the \texttt{r2\_c2} (\textit{left}) and \texttt{r0.8\_c1.5} (\textit{right}) runs at \(t\sim70\,\mathrm{Myr}\). Both runs form extended cold tails, though \texttt{r2\_c2} shows a pronounced pileup of cold gas near the source. Most of the cooling occurs along the tail interface rather than at the head. Local bow shocks are visible around the fed wake and along the coherent tail. The Brinkman core remains cold and stationary in both cases.}
    \label{fig:snapshots}
\end{figure*}

Figure~\ref{fig:snapshots} compares snapshots of the \texttt{r0.8\_c1.5} and \texttt{r2\_c2} runs at \(t\sim70\,\mathrm{Myr}\). Both develop long cold tails, and in both cases most of the cooling occurs along the cold--hot interface in the tail rather than at the head. The larger-core/shell run \texttt{r2\_c2} shows a pronounced pileup of cold gas near the source, whereas the fiducial \texttt{r0.8\_c1.5} run produces a smoother coherent tail. Our fiducial \(n_{\rm H,CGM}=5\times10^{-3}\,\mathrm{cm^{-3}}\) lies close to the minimum ambient density for forming a coherent tail in this setup; at lower densities the tail fragments. This threshold is likely sensitive to the unresolved near-head physics, so it should not be overinterpreted as a precise CGM density constraint. By contrast, in an adiabatic control run no coherent cold tail forms at all: the cold gas is rapidly ablated and mixed into the hot flow. Radiative cooling is therefore dynamically essential for producing the observed long-lived cold tail.

\begin{figure}
    \centering
    \includegraphics[width=\linewidth]{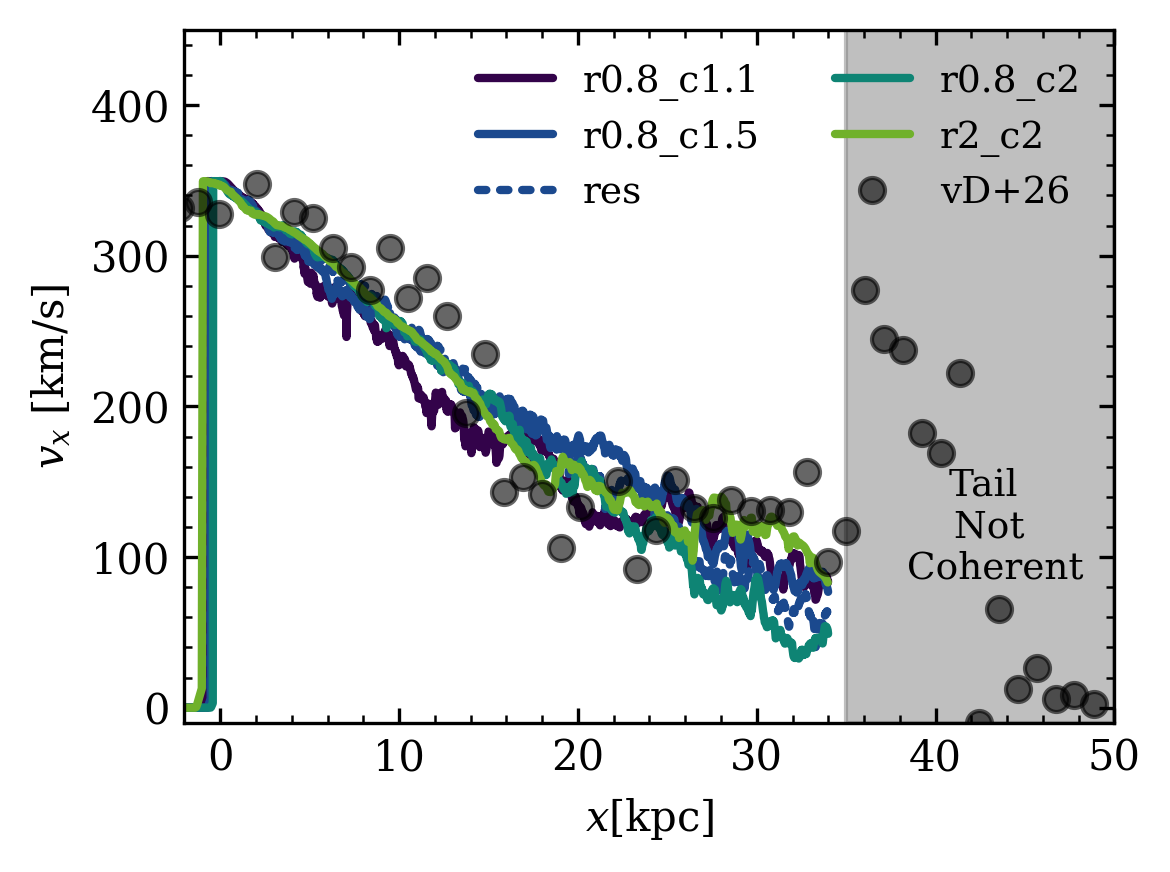}
    \caption{Projected downstream velocity profiles of the cold gas in the simulation suite, plotted as a function of distance from the start of the modeled region, i.e. \(12\,\mathrm{kpc}\) downstream of the RBH-1 head. Black circles labeled ``vD+26'' denote the H\(\alpha\) velocity measurements from \citet{vanDokkum2026} after excluding the initial \(\sim 12\,\mathrm{kpc}\) near-head region. Only runs that form coherent tails are shown. The dark gray region marks where the simulated tail is no longer coherent.}
    \label{fig:velocity_profile}
\end{figure}

Figure~\ref{fig:velocity_profile} shows the projected downstream velocity profile of the cold gas along the coherent \(\sim35\,\mathrm{kpc}\) tail, compared with the H\(\alpha\)+[O III] velocity profile measured by \citet{vanDokkum2026}. Simulation velocities are projected by \(\sin i=0.5\) to match the observed inclination. All runs that form coherent tails show good agreement with the observed \(\sim200\,\mathrm{km\,s^{-1}}\) decline across the downstream tail. As seen in Fig.~\ref{fig:velocity_profile}, the downstream velocity profile is largely insensitive to the core radius and shell thickness, with all coherent-tail runs converging to similar slopes. The dark shaded region marks where the simulated tail is no longer coherent and the cold gas is beginning to leave the computational domain; this corresponds to the observed region where no coherent tail is seen. The higher-resolution \texttt{res} run is nearly indistinguishable from the fiducial \texttt{r0.8\_c1.5} case in Figure~\ref{fig:velocity_profile}, indicating that the tail kinematics are numerically converged at the fiducial resolution.

\begin{figure}
    \centering
    \includegraphics[width=\linewidth]{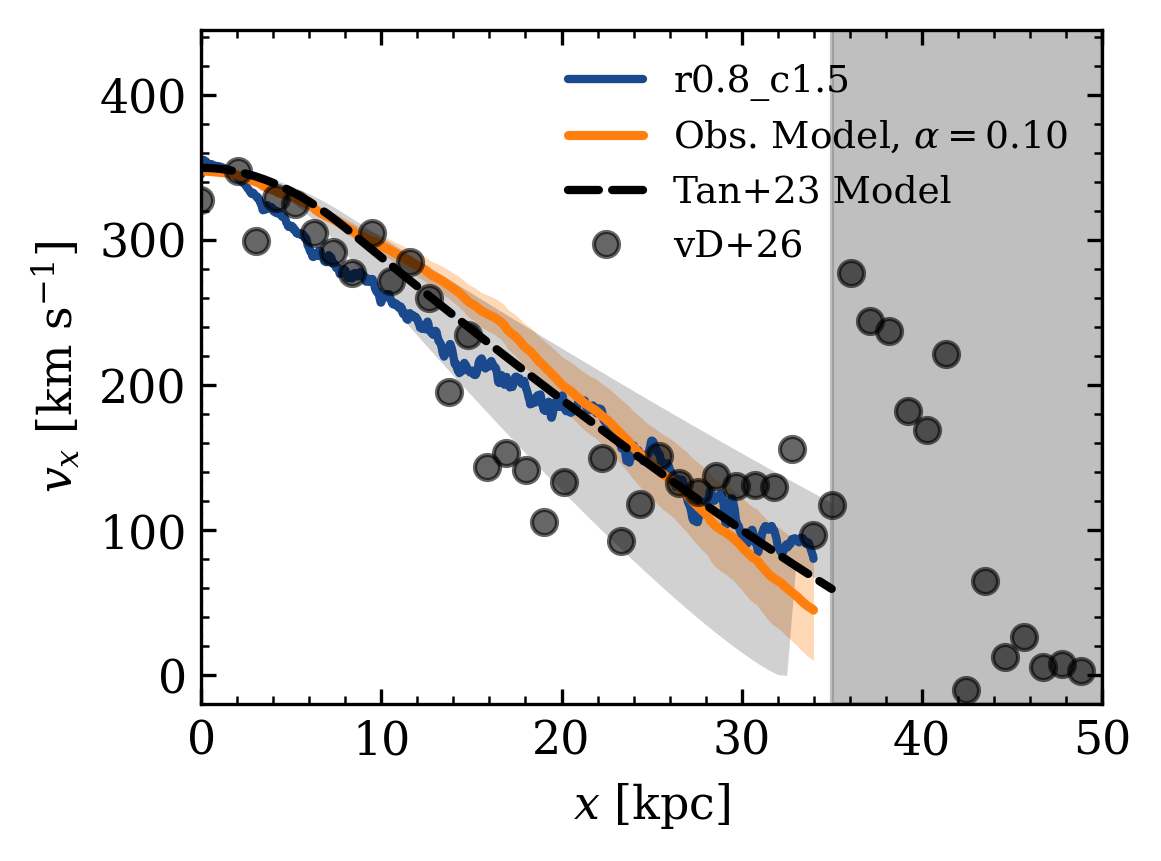}
    \caption{Velocity profile of the fiducial \texttt{r0.8\_c1.5} run compared with the two closures for the same braking law and the \citet{vanDokkum2026} data. The orange curve shows the luminosity-based closure from Equation~\ref{eqn:v_obs_pred}; the shaded band shows the \(1\sigma\) scatter associated with the inferred \(\alpha=v_{\rm src}/v\). The black dashed line shows the mixing-layer closure from Equations~\ref{eqn:velocity_equation} and \ref{eqn:tgrow_equation}, with the gray band spanning \(f_A=0.3\)--0.6.}
    \label{fig:v_observer_pred}
\end{figure}

We next compare the two closures for the same braking law introduced in \S~\ref{sec:analytic}: a luminosity-based closure using Equation~\ref{eqn:v_obs_pred}, and a mixing-layer closure using Equations~\ref{eqn:velocity_equation} and \ref{eqn:tgrow_equation}. Figure~\ref{fig:v_observer_pred} shows both for the fiducial \texttt{r0.8\_c1.5} run. The luminosity-based closure reproduces the simulated velocity profile well; allowing a small source-velocity correction, \(\alpha\simeq0.1\), modestly improves the match. The small non-zero value should be interpreted as an effective source-velocity correction: although the ambient CGM is initially at rest, gas in the shear/cooling layer can acquire some bulk motion before joining the cold tail. In practice, $\alpha\simeq 0.1$ is small and mainly rescales the normalization of the braking law. The dashed black line shows the mixing-layer closure for \(r=0.8\,\mathrm{kpc}\), \(\chi=100\), \(t_{\rm cool}=0.003\,\mathrm{Myr}\) (for \(n_{\rm H,bkg}=5\times10^{-3}\,\mathrm{cm^{-3}}\)), \(f_{\rm kh}=1\), and \(f_A=0.5\), with the gray band showing \(f_A=0.3\)--0.6. Along much of the coherent tail, both \(dL_{\rm bol}/dx\) and \(\lambda(x)\) are approximately constant, so the luminosity-based closure implies an approximately constant local \(t_{\rm grow}\) and hence an approximately linear velocity profile. The agreement between the two closures and the simulation therefore shows that the observed downstream deceleration is consistent with cooling-induced mass loading. For RBH-1, the most direct route would be spatially resolved spectroscopy along the downstream tail: in principle, recombination and forbidden-line emission could be converted into a cooling-rate profile, while density-sensitive diagnostics together with the observed tail width could constrain $\lambda(x)$. This has not yet been done: existing H$\alpha$ and [O~III] data are concentrated near the head rather than farther down the wake, and the line emission there is complicated by contributions from both star formation and localized shocks.

\begin{figure}
    \centering
    \includegraphics[width=\linewidth]{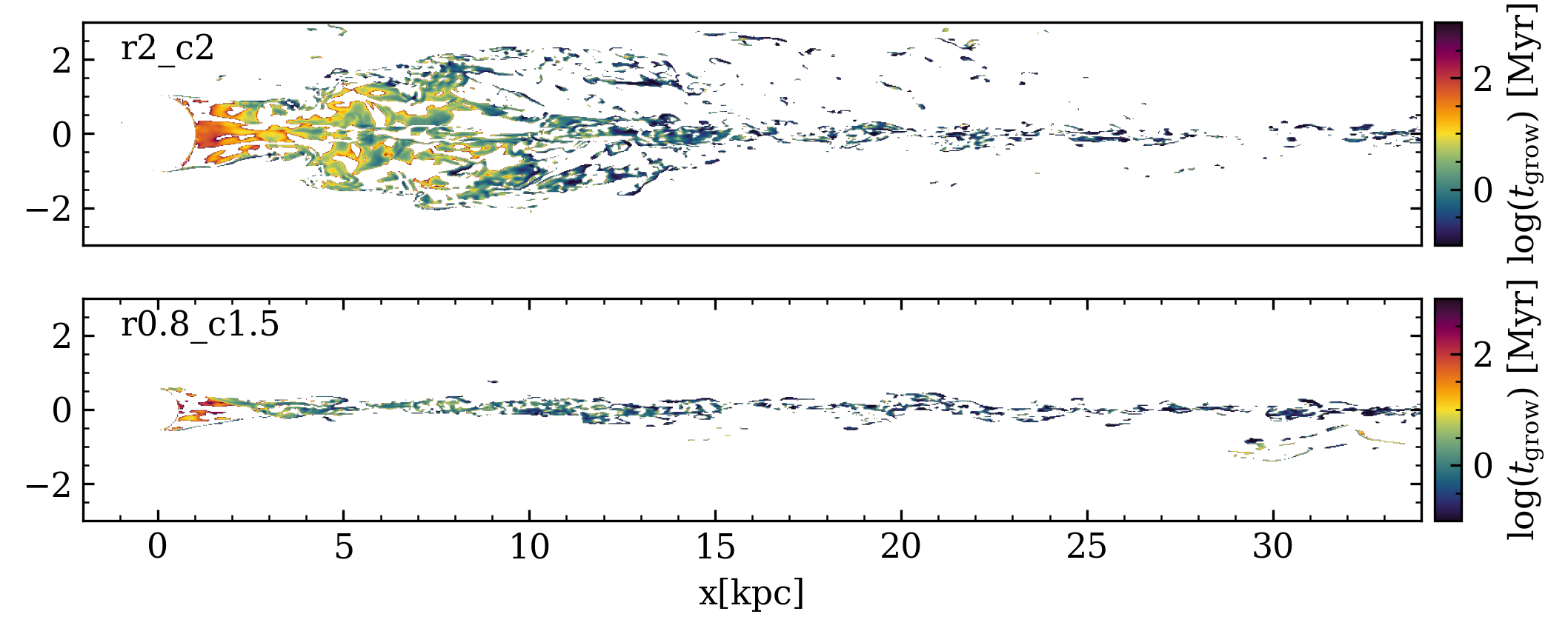}
    \caption{Slices of \(t_{\rm grow}\) for the \texttt{r2\_c2} and \texttt{r0.8\_c1.5} runs at the same time as Figure~\ref{fig:snapshots}. Near the source \(t_{\rm grow}\gg70\,\mathrm{Myr}\), while along the tail \(t_{\rm grow}\ll70\,\mathrm{Myr}\). There is little cooling directly at the head, as expected.}
    \label{fig:tgrow}
\end{figure}

Some runs, especially \texttt{r2\_c2}, show only modest net growth in total cold mass by \(t\sim70\,\mathrm{Myr}\). This does not imply weak local braking: net mass growth reflects the competition between cooling-driven gain and downstream mass loss. To isolate the local growth time, we estimate
\begin{equation}
    t_{\rm grow}=\frac{\rho}{\dot{\rho}_{\rm cold}-\dot{\rho}_{\rm adv}},
\end{equation}
where \(\dot{\rho}_{\rm adv}=-\nabla\cdot(\rho_{\rm cold}\mathbf{v})\) is the advection contribution measured from consecutive snapshots. The resulting maps are shown in Figure~\ref{fig:tgrow}. While the globally averaged growth time can be long, the local tail growth time is short, \(t_{\rm grow,tail}\ll t_{\rm lifetime}\), in both the \texttt{r2\_c2} and \texttt{r0.8\_c1.5} runs. Thus, even when the net cold mass changes little, cooling remains efficient along the tail and accretion-induced drag remains dynamically important.

At \(x\gtrsim15\,\mathrm{kpc}\) and \(t\sim t_{\rm lifetime}\), the radiatively inferred mass-addition rate exceeds the directly measured cold-mass flux growth because cold gas begins to be lost from the tail in this region. Since this lost material has \(v_{\rm loss}\sim v_{\rm tail}\), it has little effect on the deceleration, consistent with neglecting the corresponding loss term in Equation~\ref{eqn:momentum_flux}. The onset of this loss coincides with the break seen in both the simulated and observed velocity profiles, where the local advection time becomes comparable to the tail lifetime. In the simulations, this break moves downstream as the tail ages, as expected if it is set by the local advection time.

\section{Discussion}
\label{sec:discussion}

Radiative turbulent mixing layers are widely invoked to explain the survival and growth of cold gas in hot astrophysical flows, but quantitative dynamical tests have been rare. In this respect, RBH-1 is unusually valuable: once a coherent cold tail exists, its downstream deceleration can be compared directly to analytic expectations and simulations. Our main result is that the observed velocity gradient in the coherent downstream tail is well reproduced by cooling-induced mass loading, and that the inferred deceleration is consistent with both the mixing-layer and luminosity-based closures. The latter also suggests a concrete observational test: spatially resolved spectroscopy of the tail could determine whether the measured cooling luminosity and cold-gas column are sufficient to account for the observed braking.

One important question we do not answer is the origin of the cold gas near the head of RBH-1. A colder ambient CGM, as proposed by \citet{ogiya2024}, appears less consistent with the observed tail kinematics, which show both substantial downstream motion and a coherent velocity gradient. Stripping from the immediate head region is also disfavored, since the deprojected velocity at the start of the tail (\(\sim 600\,\mathrm{km\,s^{-1}}\)) is significantly below the black-hole velocity (\(\sim 950\,\mathrm{km\,s^{-1}}\)). Bow-shock-induced condensation remains plausible, but in our test simulations cold clumps form along the bow shock only for unrealistically high ambient densities, and do not fall back to form a coherent tail. The origin of this cold gas, then, remains a mystery. We therefore interpret the first \(\sim 10\)--\(12\,\mathrm{kpc}\) as a distinct formation regime that lies outside the scope of this Letter. Our results apply to the downstream tail, provided that the head physics supplies a coherent cold structure with approximately the observed initial conditions at the start of the modeled region.

While CGM magnetic fields are likely weak, they can be amplified by magnetic draping and compression during cooling. In principle, magnetic tension could suppress mixing \citep{ji19,zhao23}, while magnetic stresses could enhance momentum transfer across the interface \citep{dursi08,mccourt2015}. Nonetheless, previous MHD simulations have generally found modest effects on mass growth and entrainment, with cold gas kinematics broadly similar to the hydrodynamic case \citep{gronke2020,Kaul2025}. In one aligned-field test with \(\beta_0=100\), representative of plausible CGM conditions, we likewise find little change in the velocity profile. Stronger fields, different geometries, or anisotropic conduction could still modify the efficacy of cooling-induced braking, and remain worth exploring in future work.

\section{Conclusions}
\label{sec:conclusions}

We have used analytic estimates and 3D hydrodynamical simulations with radiative cooling to test whether the coherent downstream tail of RBH-1 can be understood as a consequence of cooling-induced entrainment. While conventional ram-pressure drag is far too weak to explain the observed braking, we find that accretion-induced drag from radiative mixing layers reproduces the \(\sim 200\,\mathrm{km\,s^{-1}}\) velocity decline along the coherent tail. Radiative cooling is dynamically essential: in an adiabatic control run, no coherent cold tail forms. We also show that the observed deceleration can be related to the cooling luminosity and cold-gas column along the tail, providing a direct route for future observational tests. Although the origin of the cold gas in the near-head region remains uncertain, the downstream tail of RBH-1 already provides one of the clearest dynamical tests to date of cooling-induced entrainment in an astrophysical system.

\section*{Acknowledgements}
We thank Pieter van Dokkum, Greg Bryan, and Zirui Chen for interesting discussions. We acknowledge NSF grant AST240752 for support. The simulations in this paper were performed on the Stampede3 computing cluster, supported by the NSF ACCESS Grant PHY240194.

\section*{Data Availability}

The data underlying this article will be shared upon reasonable
request to the corresponding author.



\bibliographystyle{mnras}
\bibliography{refs,master_references} % if your bibtex file is called example.bib

@article{zhao23,
	adsurl = {https://ui.adsabs.harvard.edu/abs/2023MNRAS.526.4245Z},
	archiveprefix = {arXiv},
	author = {{Zhao}, Xihui and {Bai}, Xue-Ning},
	doi = {10.1093/mnras/stad3011},
	eprint = {2307.12355},
	journal = {\mnras},
	month = dec,
	number = {3},
	pages = {4245-4261},
	primaryclass = {astro-ph.GA},
	title = {{Simulations of weakly magnetized turbulent mixing layers}},
	volume = {526},
	year = 2023}

@article{sharma25,
	adsurl = {https://ui.adsabs.harvard.edu/abs/2025arXiv250903802S},
	archiveprefix = {arXiv},
	author = {{Sharma}, Prateek and {Kumar}, Arnav and {Datta}, Dipayan and {Babul}, Arif and {Das}, Rishita and {Aditya}, Konduri},
	doi = {10.48550/arXiv.2509.03802},
	eid = {arXiv:2509.03802},
	eprint = {2509.03802},
	journal = {arXiv e-prints},
	month = sep,
	pages = {arXiv:2509.03802},
	primaryclass = {astro-ph.GA},
	title = {{Universal Structure of Turbulent Radiative Mixing Layers}},
	year = 2025}

@article{ji19,
	adsurl = {https://ui.adsabs.harvard.edu/abs/2019MNRAS.487..737J},
	archiveprefix = {arXiv},
	author = {{Ji}, Suoqing and {Oh}, S. Peng and {Masterson}, Phillip},
	doi = {10.1093/mnras/stz1248},
	eprint = {1809.09101},
	journal = {\mnras},
	month = {Jul},
	number = {1},
	pages = {737-754},
	primaryclass = {astro-ph.GA},
	title = {{Simulations of radiative turbulent mixing layers}},
	volume = {487},
	year = {2019}}

@article{kwak10,
	adsurl = {http://adsabs.harvard.edu/abs/2010ApJ...719..523K},
	archiveprefix = {arXiv},
	author = {{Kwak}, K. and {Shelton}, R.~L.},
	doi = {10.1088/0004-637X/719/1/523},
	eprint = {1006.3811},
	journal = {\apj},
	month = aug,
	pages = {523-539},
	primaryclass = {astro-ph.GA},
	title = {{Numerical Study of Turbulent Mixing Layers with Non-equilibrium Ionization Calculations}},
	volume = 719,
	year = 2010}

@article{dursi08,
	adsurl = {http://adsabs.harvard.edu/abs/2008ApJ...677..993D},
	archiveprefix = {arXiv},
	author = {{Dursi}, L.~J. and {Pfrommer}, C.},
	doi = {10.1086/529371},
	eprint = {0711.0213},
	journal = {\apj},
	month = apr,
	pages = {993-1018},
	title = {{Draping of Cluster Magnetic Fields over Bullets and Bubbles-Morphology and Dynamic Effects}},
	volume = 677,
	year = 2008}

@article{begelman90,
	adsurl = {http://adsabs.harvard.edu/abs/1990MNRAS.244P..26B},
	author = {{Begelman}, M.~C. and {Fabian}, A.~C.},
	journal = {\mnras},
	month = may,
	pages = {26P-29P},
	title = {{Turbulent mixing layers in the interstellar and intracluster medium}},
	volume = 244,
	year = 1990}

@ARTICLE{vanDokkum2026,
       author = {{van Dokkum}, Pieter and {Jennings}, Connor and {Pasha}, Imad and {Conroy}, Charlie and {Kaul}, Ish and {Abraham}, Roberto and {Danieli}, Shany and {Romanowsky}, Aaron J. and {Tremblay}, Grant},
        title = "{JWST Confirmation of a Runaway Supermassive Black Hole via Its Supersonic Bow Shock}",
      journal = {\apjl},
         year = 2026,
        month = feb,
       volume = {998},
       number = {1},
          eid = {L27},
        pages = {L27},
          doi = {10.3847/2041-8213/ae3d0e},
archivePrefix = {arXiv},
       eprint = {2512.04166},
 primaryClass = {astro-ph.GA},
       adsurl = {https://ui.adsabs.harvard.edu/abs/2026ApJ...998L..27V}
}

@ARTICLE{vanDokkum2023,
       author = {{van Dokkum}, Pieter and {Pasha}, Imad and {Buzzo}, Maria Luisa and {LaMassa}, Stephanie and {Shen}, Zili and {Keim}, Michael A. and {Abraham}, Roberto and {Conroy}, Charlie and {Danieli}, Shany and {Mitra}, Kaustav and {Nagai}, Daisuke and {Natarajan}, Priyamvada and {Romanowsky}, Aaron J. and {Tremblay}, Grant and {Urry}, C. Megan and {van den Bosch}, Frank C.},
        title = "{A Candidate Runaway Supermassive Black Hole Identified by Shocks and Star Formation in its Wake}",
      journal = {\apjl},
         year = 2023,
        month = apr,
       volume = {946},
       number = {2},
          eid = {L50},
        pages = {L50},
          doi = {10.3847/2041-8213/acba86},
archivePrefix = {arXiv},
       eprint = {2302.04888},
 primaryClass = {astro-ph.GA},
       adsurl = {https://ui.adsabs.harvard.edu/abs/2023ApJ...946L..50V}
}

@ARTICLE{Tan2023,
       author = {{Tan}, Brent and {Oh}, S. Peng and {Gronke}, Max},
        title = "{Cloudy with a chance of rain: accretion braking of cold clouds}",
      journal = {\mnras},
         year = 2023,
        month = apr,
       volume = {520},
       number = {2},
        pages = {2571-2592},
          doi = {10.1093/mnras/stad236},
archivePrefix = {arXiv},
       eprint = {2210.06493},
 primaryClass = {astro-ph.GA},
       adsurl = {https://ui.adsabs.harvard.edu/abs/2023MNRAS.520.2571T}
}

@ARTICLE{Kaul2025,
       author = {{Kaul}, Ish and {Tan}, Brent and {Oh}, S. Peng and {Mandelker}, Nir},
        title = "{Tales of tension: magnetized infalling cold clouds and streams in the CGM}",
      journal = {\mnras},
         year = 2025,
        month = jun,
       volume = {539},
       number = {4},
        pages = {3669-3696},
          doi = {10.1093/mnras/staf706},
archivePrefix = {arXiv},
       eprint = {2502.17549},
 primaryClass = {astro-ph.GA},
       adsurl = {https://ui.adsabs.harvard.edu/abs/2025MNRAS.539.3669K}
}

@ARTICLE{Gronke2018,
       author = {{Gronke}, Max and {Oh}, S. Peng},
        title = "{The growth and entrainment of cold gas in a hot wind}",
      journal = {\mnras},
         year = 2018,
        month = oct,
       volume = {480},
       number = {1},
        pages = {L111-L115},
          doi = {10.1093/mnrasl/sly131},
archivePrefix = {arXiv},
       eprint = {1806.02728},
 primaryClass = {astro-ph.GA},
       adsurl = {https://ui.adsabs.harvard.edu/abs/2018MNRAS.480L.111G}
}

@ARTICLE{Chen2024,
       author = {{Chen}, Zirui and {Oh}, S. Peng},
        title = "{The survival and entrainment of molecules and dust in galactic winds}",
      journal = {\mnras},
         year = 2024,
        month = jun,
       volume = {530},
       number = {4},
        pages = {4032-4057},
          doi = {10.1093/mnras/stae1113},
archivePrefix = {arXiv},
       eprint = {2311.04275},
 primaryClass = {astro-ph.GA},
       adsurl = {https://ui.adsabs.harvard.edu/abs/2024MNRAS.530.4032C}
}

@ARTICLE{Li2020,
       author = {{Li}, Zhihui and {Hopkins}, Philip F. and {Squire}, Jonathan and {Hummels}, Cameron},
        title = "{On the survival of cool clouds in the circumgalactic medium}",
      journal = {\mnras},
         year = 2020,
        month = feb,
       volume = {492},
       number = {2},
        pages = {1841-1854},
          doi = {10.1093/mnras/stz3567},
archivePrefix = {arXiv},
       eprint = {1909.02632},
 primaryClass = {astro-ph.GA},
       adsurl = {https://ui.adsabs.harvard.edu/abs/2020MNRAS.492.1841L}
}

@ARTICLE{Yang2023,
       author = {{Yang}, Yanhui and {Ji}, Suoqing},
        title = "{Radiative turbulent mixing layers at high Mach numbers}",
      journal = {\mnras},
         year = 2023,
        month = apr,
       volume = {520},
       number = {2},
        pages = {2148-2162},
          doi = {10.1093/mnras/stad264},
archivePrefix = {arXiv},
       eprint = {2205.15336},
 primaryClass = {astro-ph.GA},
       adsurl = {https://ui.adsabs.harvard.edu/abs/2023MNRAS.520.2148Y}
}

@ARTICLE{Tan2021,
       author = {{Tan}, Brent and {Oh}, S. Peng and {Gronke}, Max},
        title = "{Radiative mixing layers: insights from turbulent combustion}",
      journal = {\mnras},
         year = 2021,
        month = apr,
       volume = {502},
       number = {3},
        pages = {3179-3199},
          doi = {10.1093/mnras/stab053},
archivePrefix = {arXiv},
       eprint = {2008.12302},
 primaryClass = {astro-ph.GA},
       adsurl = {https://ui.adsabs.harvard.edu/abs/2021MNRAS.502.3179T}
}

@ARTICLE{fielding2020,
       author = {{Fielding}, Drummond B. and {Ostriker}, Eve C. and {Bryan}, Greg L. and {Jermyn}, Adam S.},
        title = "{Multiphase Gas and the Fractal Nature of Radiative Turbulent Mixing Layers}",
      journal = {\apjl},
         year = 2020,
        month = may,
       volume = {894},
       number = {2},
          eid = {L24},
        pages = {L24},
          doi = {10.3847/2041-8213/ab8d2c},
archivePrefix = {arXiv},
       eprint = {2003.08390},
 primaryClass = {astro-ph.GA},
       adsurl = {https://ui.adsabs.harvard.edu/abs/2020ApJ...894L..24F}
}

@ARTICLE{chen2023,
       author = {{Chen}, Zirui and {Fielding}, Drummond B. and {Bryan}, Greg L.},
        title = "{The Anatomy of a Turbulent Radiative Mixing Layer: Insights from an Analytic Model with Turbulent Conduction and Viscosity}",
      journal = {\apj},
         year = 2023,
        month = jun,
       volume = {950},
       number = {2},
          eid = {91},
        pages = {91},
          doi = {10.3847/1538-4357/acc73f},
archivePrefix = {arXiv},
       eprint = {2211.01395},
 primaryClass = {astro-ph.GA},
       adsurl = {https://ui.adsabs.harvard.edu/abs/2023ApJ...950...91C}
}

@ARTICLE{hidalgo-pineda2024,
       author = {{Hidalgo-Pineda}, Fernando and {Farber}, Ryan Jeffrey and {Gronke}, Max},
        title = "{Better together: the complex interplay between radiative cooling and magnetic draping}",
      journal = {\mnras},
         year = 2024,
        month = jan,
       volume = {527},
       number = {1},
        pages = {135-149},
          doi = {10.1093/mnras/stad3069},
archivePrefix = {arXiv},
       eprint = {2304.09897},
 primaryClass = {astro-ph.GA},
       adsurl = {https://ui.adsabs.harvard.edu/abs/2024MNRAS.527..135H}
}

@ARTICLE{marin-gilabert2025,
       author = {{Marin-Gilabert}, Tirso and {Gronke}, Max and {Oh}, S. Peng},
        title = "{The (Limited) Effect of Viscosity in Multiphase Turbulent Mixing}",
      journal = {arXiv e-prints},
         year = 2025,
        month = apr,
          eid = {arXiv:2504.15345},
        pages = {arXiv:2504.15345},
          doi = {10.48550/arXiv.2504.15345},
archivePrefix = {arXiv},
       eprint = {2504.15345},
 primaryClass = {astro-ph.GA},
       adsurl = {https://ui.adsabs.harvard.edu/abs/2025arXiv250415345M}
}

@ARTICLE{gronke2020,
       author = {{Gronke}, Max and {Oh}, S. Peng},
        title = "{How cold gas continuously entrains mass and momentum from a hot wind}",
      journal = {\mnras},
         year = 2020,
        month = feb,
       volume = {492},
       number = {2},
        pages = {1970-1990},
          doi = {10.1093/mnras/stz3332},
archivePrefix = {arXiv},
       eprint = {1907.04771},
 primaryClass = {astro-ph.GA},
       adsurl = {https://ui.adsabs.harvard.edu/abs/2020MNRAS.492.1970G}
}

@ARTICLE{jennings2023,
       author = {{Jennings}, Fred and {Beckmann}, Ricarda S. and {Sijacki}, Debora and {Dubois}, Yohan},
        title = "{Shattering and growth of cold clouds in galaxy clusters: the role of radiative cooling, magnetic fields, and thermal conduction}",
      journal = {\mnras},
         year = 2023,
        month = feb,
       volume = {518},
       number = {4},
        pages = {5215-5235},
          doi = {10.1093/mnras/stac3426},
archivePrefix = {arXiv},
       eprint = {2211.09183},
 primaryClass = {astro-ph.GA},
       adsurl = {https://ui.adsabs.harvard.edu/abs/2023MNRAS.518.5215J}
}

@ARTICLE{stone2020,
       author = {{Stone}, James M. and {Tomida}, Kengo and {White}, Christopher J. and {Felker}, Kyle G.},
        title = "{The Athena++ Adaptive Mesh Refinement Framework: Design and Magnetohydrodynamic Solvers}",
      journal = {\apjs},
         year = 2020,
        month = jul,
       volume = {249},
       number = {1},
          eid = {4},
        pages = {4},
          doi = {10.3847/1538-4365/ab929b},
archivePrefix = {arXiv},
       eprint = {2005.06651},
 primaryClass = {astro-ph.IM},
       adsurl = {https://ui.adsabs.harvard.edu/abs/2020ApJS..249....4S}
}

@ARTICLE{gnat2007,
       author = {{Gnat}, Orly and {Sternberg}, Amiel},
        title = "{Time-dependent Ionization in Radiatively Cooling Gas}",
      journal = {\apjs},
         year = 2007,
        month = feb,
       volume = {168},
       number = {2},
        pages = {213-230},
          doi = {10.1086/509786},
archivePrefix = {arXiv},
       eprint = {astro-ph/0608181},
 primaryClass = {astro-ph},
       adsurl = {https://ui.adsabs.harvard.edu/abs/2007ApJS..168..213G}
}

@ARTICLE{liu2007,
       author = {{Liu}, Qianlong and {Vasilyev}, Oleg V.},
        title = "{A Brinkman penalization method for compressible flows in complex geometries}",
      journal = {Journal of Computational Physics},
         year = 2007,
        month = dec,
       volume = {227},
       number = {2},
        pages = {946-966},
          doi = {10.1016/j.jcp.2007.07.037},
       adsurl = {https://ui.adsabs.harvard.edu/abs/2007JCoPh.227..946L}
}

@ARTICLE{westmeier2018,
       author = {{Westmeier}, Tobias},
        title = "{A new all-sky map of Galactic high-velocity clouds from the 21-cm HI4PI survey}",
      journal = {\mnras},
         year = 2018,
        month = feb,
       volume = {474},
       number = {1},
        pages = {289-299},
          doi = {10.1093/mnras/stx2757},
archivePrefix = {arXiv},
       eprint = {1712.00909},
 primaryClass = {astro-ph.GA},
       adsurl = {https://ui.adsabs.harvard.edu/abs/2018MNRAS.474..289W}
}

@ARTICLE{Fox2005,
       author = {{Fox}, Andrew J. and {Wakker}, Bart P. and {Savage}, Blair D. and {Tripp}, Todd M. and {Sembach}, Kenneth R. and {Bland-Hawthorn}, Joss},
        title = "{Multiphase High-Velocity Clouds toward HE 0226-4110 and PG 0953+414}",
      journal = {\apj},
         year = 2005,
        month = sep,
       volume = {630},
       number = {1},
        pages = {332-354},
          doi = {10.1086/431915},
archivePrefix = {arXiv},
       eprint = {astro-ph/0505299},
 primaryClass = {astro-ph},
       adsurl = {https://ui.adsabs.harvard.edu/abs/2005ApJ...630..332F}
}

@ARTICLE{Lim2008,
       author = {{Lim}, Jeremy and {Ao}, YiPing and {Dinh-V-Trung}},
        title = "{Radially Inflowing Molecular Gas in NGC 1275 Deposited by an X-Ray Cooling Flow in the Perseus Cluster}",
      journal = {\apj},
         year = 2008,
        month = jan,
       volume = {672},
       number = {1},
        pages = {252-265},
          doi = {10.1086/523664},
archivePrefix = {arXiv},
       eprint = {0712.2979},
 primaryClass = {astro-ph},
       adsurl = {https://ui.adsabs.harvard.edu/abs/2008ApJ...672..252L}
}

@ARTICLE{ogiya2024,
       author = {{Ogiya}, Go and {Nagai}, Daisuke},
        title = "{Formation of dense filaments induced by runaway supermassive black holes}",
      journal = {\mnras},
         year = 2024,
        month = jan,
       volume = {527},
       number = {3},
        pages = {5503-5513},
          doi = {10.1093/mnras/stad3469},
archivePrefix = {arXiv},
       eprint = {2309.09031},
 primaryClass = {astro-ph.GA},
       adsurl = {https://ui.adsabs.harvard.edu/abs/2024MNRAS.527.5503O}
}

@ARTICLE{faucher-giguere2023,
       author = {{Faucher-Gigu{\`e}re}, Claude-Andr{\'e} and {Oh}, S. Peng},
        title = "{Key Physical Processes in the Circumgalactic Medium}",
      journal = {\araa},
         year = 2023,
        month = aug,
       volume = {61},
        pages = {131-195},
          doi = {10.1146/annurev-astro-052920-125203},
archivePrefix = {arXiv},
       eprint = {2301.10253},
 primaryClass = {astro-ph.GA},
       adsurl = {https://ui.adsabs.harvard.edu/abs/2023ARA&A..61..131F}
}

@ARTICLE{mandelker2020,
       author = {{Mandelker}, Nir and {Nagai}, Daisuke and {Aung}, Han and {Dekel}, Avishai and {Birnboim}, Yuval and {van den Bosch}, Frank C.},
        title = "{Instability of supersonic cold streams feeding galaxies - IV. Survival of radiatively cooling streams}",
      journal = {\mnras},
         year = 2020,
        month = may,
       volume = {494},
       number = {2},
        pages = {2641-2663},
          doi = {10.1093/mnras/staa812},
archivePrefix = {arXiv},
       eprint = {1910.05344},
 primaryClass = {astro-ph.GA},
       adsurl = {https://ui.adsabs.harvard.edu/abs/2020MNRAS.494.2641M}
}

@ARTICLE{mccourt2015,
       author = {{McCourt}, Michael and {O'Leary}, Ryan M. and {Madigan}, Ann-Marie and {Quataert}, Eliot},
        title = "{Magnetized gas clouds can survive acceleration by a hot wind}",
      journal = {\mnras},
         year = 2015,
        month = may,
       volume = {449},
       number = {1},
        pages = {2-7},
          doi = {10.1093/mnras/stv355},
archivePrefix = {arXiv},
       eprint = {1409.6719},
 primaryClass = {astro-ph.GA},
       adsurl = {https://ui.adsabs.harvard.edu/abs/2015MNRAS.449....2M}
}

%\begin{thebibliography}{99}
%\end{thebibliography}

%%%%%%%%%%%%%%%%%%%%%%%%%%%%%%%%%%%%%%%%%%%%%%%%%%

%%%%%%%%%%%%%%%%% APPENDICES %%%%%%%%%%%%%%%%%%%%%

%\appendix


%%%%%%%%%%%%%%%%%%%%%%%%%%%%%%%%%%%%%%%%%%%%%%%%%%

% Don't change these lines
\bsp	% typesetting comment
\label{lastpage}
\end{document}